\documentclass[conference]{IEEEtran}
\IEEEoverridecommandlockouts

\usepackage{cite}
\usepackage{amsmath,amssymb,amsfonts}
\usepackage{algorithm}
\usepackage{algpseudocode}
\usepackage{subcaption}
\usepackage{graphicx}
\usepackage{multirow}
\usepackage{textcomp}
\usepackage{bytefield}
\usepackage{comment}
\usepackage{xcolor}

\def\BibTeX{{\rm B\kern-.05em{\sc i\kern-.025em b}\kern-.08em
    T\kern-.1667em\lower.7ex\hbox{E}\kern-.125emX}}

\usepackage{float}

\usepackage[colorinlistoftodos,prependcaption,textsize=footnotesize]{todonotes}

\usepackage[most]{tcolorbox}

\newcommand{\etal}{\textit{et al}. }
\newcommand{\ie}{\textit{i}.\textit{e}., }
\newcommand{\eg}{\textit{e}.\textit{g}.\ }

\usepackage{tikz}
\usepackage{textcomp}
\usepackage{hyperref}
\usepackage{lipsum}

\newcommand\copyrighttext{%
  \footnotesize \textcopyright~  2023 IEEE.  Personal use of this material is permitted.  Permission from IEEE must be obtained for all other uses, in any current or future media, including reprinting/republishing this material for advertising or promotional purposes, creating new collective works, for resale or redistribution to servers or lists, or reuse of any copyrighted component of this work in other works. The official version can be found at
  \href{https://doi.org/10.1109/VNC57357.2023.10136285}{https://doi.org/10.1109/VNC57357.2023.10136285}}
\newcommand\copyrightnotice{%
\begin{tikzpicture}[remember picture,overlay]
\node[anchor=south,yshift=10pt] at (current page.south) {\fbox{\parbox{\dimexpr\textwidth-\fboxsep-\fboxrule\relax}{\copyrighttext}}};
\end{tikzpicture}%
}

\begin{document}

\title{On the Resilience of Machine Learning-Based IDS for Automotive Networks}

\author{\IEEEauthorblockN{Ivo Zenden, Han Wang, Alfonso Iacovazzi, Arash Vahidi, Rolf Blom, Shahid Raza}
%\author{\IEEEauthorblockN{TBD}
%\IEEEauthorblockA{%\textit{Cybersecurity Unit @ RISE} \\
 RISE Research Institutes of Sweden,
  %Isafjordsgatan 22, 164 40
  Kista, Sweden
  }

\maketitle
\copyrightnotice

\begin{abstract}

Modern automotive functions are controlled by a large number of small computers called electronic control units (ECUs). These functions span from safety-critical autonomous driving to comfort and infotainment. ECUs communicate with one another over multiple internal networks using different technologies. Some, such as Controller Area Network (CAN), are very simple and provide minimal or no security services. Machine learning techniques can be used to detect anomalous activities in such networks. However, it is necessary that these machine learning techniques are not prone to adversarial attacks.

In this paper, we investigate adversarial sample vulnerabilities in four different machine learning-based intrusion detection systems for automotive networks.
We show that adversarial samples negatively impact three of the four studied solutions. Furthermore, we analyze transferability of adversarial samples between different systems. We also investigate detection performance and the attack success rate after using adversarial samples in the training. After analyzing these results, we discuss whether current solutions are mature enough for a use in modern vehicles.

\end{abstract}


\section{Introduction}

The amount of electronics and software in a modern car has been steadily increasing. Automotive software can reach tens or even hundreds of millions of lines of code executing in a large number of ECUs \cite{IEEE_This_Car_Runs_On_Code_2009}. The ECUs are responsible for various vehicle functions such as driving and comfort and communicate over multiple in-vehicle networks. They often employ different network technologies, such as Ethernet and Controller Area Network (CAN). CAN is an automotive bus technology for communication within a vehicle \cite{Bosh_CAN_Specification_1991}. It
allows ECUs and some peripherals to communicate over a shared bus using a simple protocol.

This increase in software size, complexity, and connectivity also has led to an increased attack surface and resulted in a number of vulnerabilities \cite{Comprehensive_Experimental_Analyses_Automotive_Attack_Surfaces_2011}.
Some of the used technologies have inherent security shortcomings that may allow an attacker to gain control over parts of the vehicle and maybe even critical functions such as braking and engine control. There have been attempts to introduce some security functionalities \textit{post facto}, for example by adding integrity mechanisms to previously unprotected communication channels, or employing hardware security modules for cryptographic operations \cite{Efficient_In_Vehicle_Delayed_Data_Authentication_Compound_Message_Authentication_Code_2008, State_Of_The_Art_Embedding_Security_In_Vehicles_2007}. Unfortunately, most such solutions have either not been broadly adopted and/or provide only partial protection.

Intrusion Detection Systems (IDS), already widely used in IT systems, have also been proposed as a technology that might improve automotive security \cite{strandberg2021}. Such systems generally operate by observing system events (for example incoming network packets) and using predefined rules to detect threats or utilizing knowledge about expected behaviour to detect anomalies. Many different approaches to anomaly detection have been investigated including statistical and probabilistic models as well as machine learning (ML) in different shapes, such as
K-Nearest Neighbors (KNN), Support Vector Machine (SVM)
and Random Forest (RF) \cite{survey_data_mining_ml_methods_cyber_security_intrusion_detection_2016}. Recently, the use of artificial neural networks and Deep Learning (DL) has gained popularity \cite{Deep_Learning_Anomaly_Detection_Survey_2019}. Different DL solutions utilizing convolutional, recurrent, coding networks, etc. have been demonstrated to perform well \cite{CANIntelliIDS_2021, CANet_2020, CANTransfer_2020}. However, only few works have considered the possibility of the IDS itself to be the target of an attack. Therefore, it remains uncertain if currently available IDSs will be effective against actual adversarial attacks.

In this work we investigate security issues related to state-of-the-art ML-based CAN network IDSs, focusing on the use of adversarial samples as a means of model evasion attack.
We evaluate the performance of multiple IDS implementations with and without adversarial samples, generated by techniques such as Fast Gradient-Sign Method (FGSM) \cite{Explaining_Harnessing_Adversarial_Examples_2014}.
Empirical results show that adversarial samples generally do impact detection rates. However, retraining the ML model with adversarial samples can lead to improved performance. Furthermore, our experiments show that adversarial samples can be transferable between different models. The contributions of this paper can be summarized as follows:
\begin{itemize}
    \item We investigate the resilience and vulnerabilities of four ML-based IDSs in automotive networks against model evasion attacks performed with adversarial samples.
    \item We formulate different strategies for feature selection in the generation of adversarial samples according to three different scenarios that include DoS, fuzzy, and malfunction attacks.
    \item We perform adversarial training on the ML-based IDSs with adversarial samples and investigate performance improvement and resistance to evasion attacks.
    \item We highlight the current limitations of  ML-based IDSs, which need to be addressed in future IDSs due to security and safety considerations.
\end{itemize}

\section{Related Work}
\label{sec:relatedwork}

In this section, we present the related works on automotive IDS, model evasion attacks against such systems, and ML-based defense techniques.

\subsection{Intrusion Detection Systems for Vehicle Networks}

Recent solutions make often use of ML/DL algorithms, thanks to the high accuracy they can achieve in detecting attacks that exploit CAN protocol vulnerabilities.
For example, Tariq \etal \cite{CANTransfer_2020} combine convolutional neural networks (CNN) and Long short-term memory (LSTM) in CANTransfer, which allows to effectively model multivariate time series data. CANTransfer utilizes one-shot learning, which entails that the model can learn to detect new attacks after being trained on only a single data sample of that attack. Hanselmann \etal \cite{CANet_2020} present CANet to analyse each CAN transmitter (each sender ID) separately using a LSTM model, the result of which are combined into a "joint latent vector" which is eventually fed into an Autoencoder. This allows capturing two key system characteristics: the behaviour of each ECU by itself and its interaction with other ECUs. Javed \etal \cite{CANIntelliIDS_2021} propose CANIntelliIDS with the aim of learning the link between CAN transmitters by using a model that combines CNN and attention-based gated recurrent units (AGRU). The convolutional layers retrieve various features from the inputs, while AGRU layers learn sequential and contextual information. These existing solutions for ML-based IDS are generally good in detecting abnormal or malicious traffic on CAN buses.

\subsection{Model evasion attacks against ML-based IDSs}

Evasion attacks are of particular interest to IDS implementations as they could potentially allow an adversary to avoid detection \cite{Adversarial_Attacks_Against_Intrusion_Detction_Systems_Taxonomy_2013}. One common method for evasion attacks is to exploit the model weakness to tailor perturbations. For example, Ayub~\etal~\cite{Model_Evasion_Attack_IDS_Adversarial_ML_2020} demonstrate how the accuracy of a multi-layer perception used for intrusion detection in computer networks is reduced by at least 20 percentage points using the Jacobian-based Saliency Map Attack . Similarly, Li \etal \cite{Adversarial_Attack_Defending_System_Securing_Vehicle_Network_2021} demonstrate 98\% and 99\% success rate in evading detection from an automotive IDS for CAN signal values by creating adversarial samples with the FGSM and  Basic Iterative Method (BIM) \cite{Adversarial_Attack_Defending_System_Securing_Vehicle_Network_2021}. The work from Li \etal is similar to ours, while Li \etal look at the data in a higher protocol level. Observing from these works, we find adversarial samples are the critical threat to ML-based IDS to launch model evasion attack.

\subsection{Defenses against adversarial sample}

To increase the resilience of ML methods against adversarial samples, adversarial training is one kind of defense. In \cite{Limitations_Deep_Learning_Adversarial_Setting_2016}, training on adversarial samples reduces the success rate of such samples by 7 percentage points. Additionally, generating the new adversarial samples requires a larger perturbation size for them to be successful. Moosavi-Dezfooli \etal \cite{Deepfool_2016} propose DeepFool, which utilizes adversarial training to defend against adversarial attacks. Note however, they also show an adverse affect on the robustness if the perturbations are too large. Similarly, Grosse \etal \cite{Adversarial_Perturbations_Against_Deep_Neural_Networks_Malware_Classification_2016} look into testing different ratios of adversarial samples in the datasets and show that adding adversarial samples to the training dataset does not always increase robustness. They analyze the values of a threshold of the number of adversarial samples in the dataset that could actually affect the model performance. These defense techniques show their effectiveness on mitigating the impact brought by the adversarial samples. In this work we also  show the effectiveness of adversarial training as the defense of ML-based IDS for automotive networks.

\section{Attack Scenario and Assumptions}
\label{sec:problemmodel}
We adopt an attack model that is in accordance with the models adopted by Jiang \etal \cite{Poisoning_Evasion_DL_Vehicles} and Apruzzese~\etal~\cite{Modeling_Realistic_Attacks_NIDS}.
The considered attack scenario assumes that the attacker has compromised a single ECU in a target vehicle. The attacker can use the compromised ECU to read and inject CAN messages to learn as much as possible about the workings of the target IDS. The attacker can also observe the results generated by the IDS. Thus, they have black box knowledge about the studied IDS, but have the capability to use the IDS in the vehicle.
Therefore, it is also assumed that the attacker has no knowledge about or access to the training and test datasets used.

The attack proceeds as follows: The attacker sends several messages over the CAN and collects the corresponding IDS results. In this way they can create a dataset which is used to train an IDS model of ones choosing. The chosen model is then a white box model and as the training data for this model is known, it is possible for the attacker to create adversarial samples according to his preferred method. The adversarial samples created are then transferred onto the CAN in the compromised vehicle to fool the target IDS.

\section{Methodology}
\label{sec:approach}

\begin{figure}[!ht]
  \begin{center}
    \includegraphics[width=\linewidth]{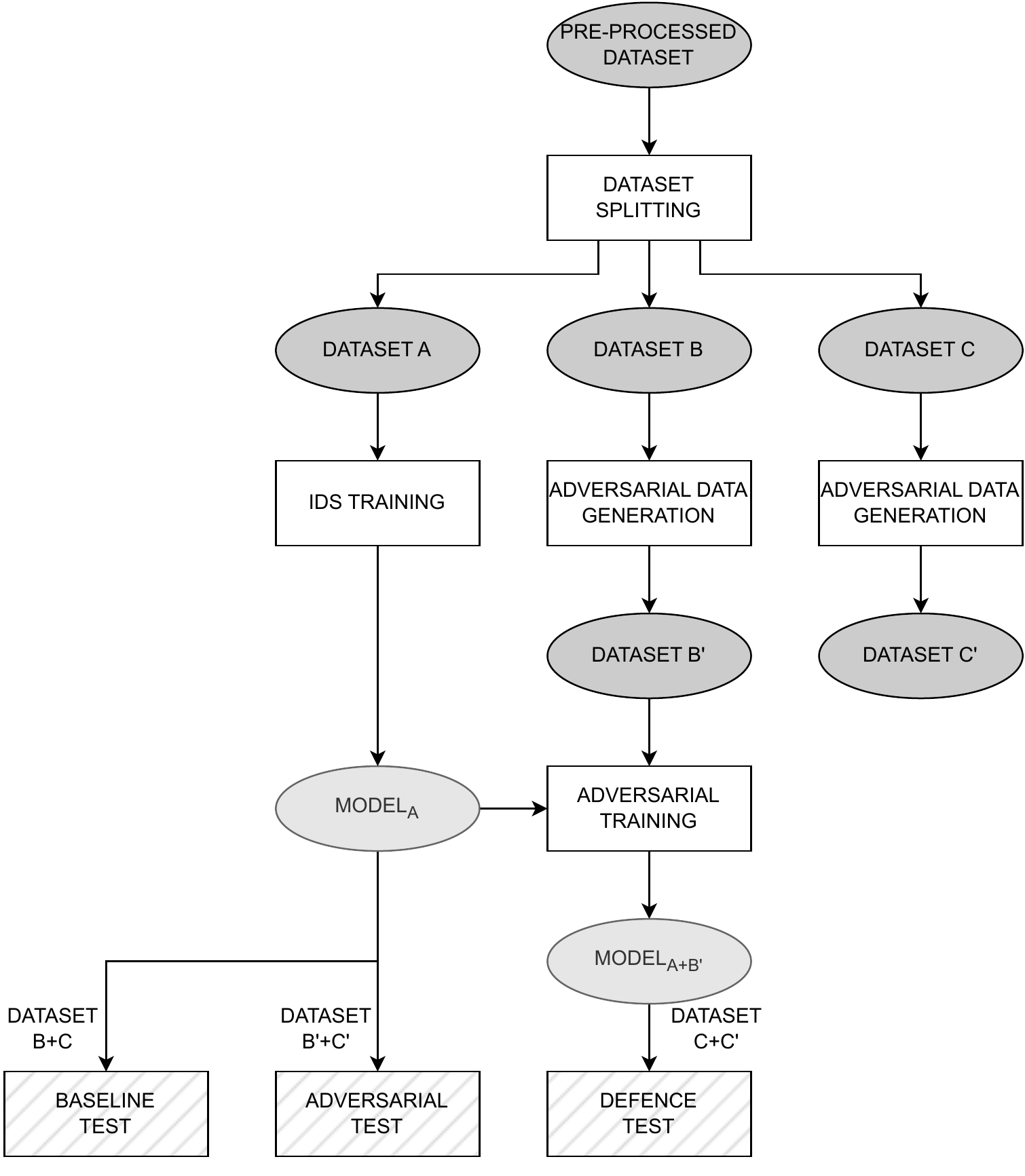}
  \end{center}
  \caption{Overview of the data flow}
  \label{fig:data-flow}
\end{figure}

Figure~\ref{fig:data-flow} shows the flow diagram followed in our approach to analyze and evaluate the resilience of ML-based IDSs. The file logs collected from automotive CAN buses are first pre-processed to extract data features and obtain a proper full dataset. The dataset is then split into three sub-sets with different purposes in the process. One subset is used to train a baseline model, the other two ones are perturbed to analyze the impact of the adversarial attack. Finally, three types of tests are performed: baseline, adversarial, and defence tests. This set of tests helps to understand if and how the performance of selected models varies when the samples are perturbed.

\subsection{Dataset description and pre-processing}

In this work, we use Survival dataset\footnote{https://ocslab.hksecurity.net/Datasets/survival-ids} \cite{Survival_dataset} as the example. Survival dataset contains several logs of CAN traffic for three different vehicles: a HYUNDAI YF Sonata 2010, a KIA compact SUV Soul 2015, and a CHEVROLET mini-compact vehicle Spark 2015. For each of these vehicles, four different log files are recorded. One with only normal CAN traffic, one with DoS attacks, one with Fuzzy attacks and the final one with Malfunction attacks. We mainly focus on the data from HYUNDAI YF Sonata 2010.

\begin{figure}[!ht]
    \includegraphics[scale=0.083]{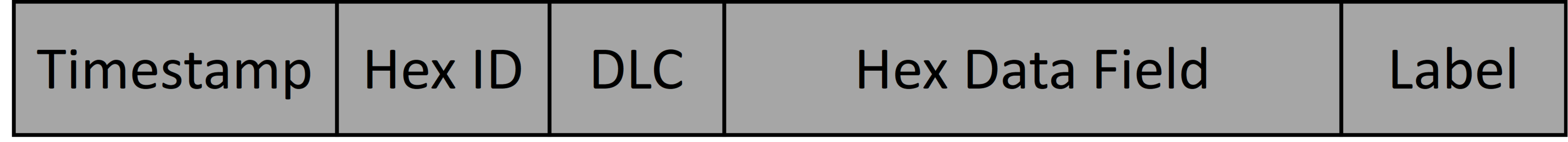}
  \caption{CAN log}
  \label{fig:can-log}
\end{figure}

Typically, CAN traffic gathered from vehicles is made available in the form of log files; the format of the records in the log files is provided in Figure~\ref{fig:can-log}. The first field is a timestamp for when the message is observed. Then, the record includes the message ID in hexadecimal format which is followed by the Data Length Code (DLC). Next, the data field is the actual content of the message which may contain up to 8 data bytes recorded in hexadecimal format. Finally, the label field contains either R or T signalling normal and attack messages, respectively. %

To transform the data and make them suited for the input of the ML algorithms, several pre-processing steps are taken. (i) All the hexadecimal values are re-converted into
binary values. %
(ii) %
The new ``delta Time'' (dTIME) feature substitutes the timestamp.
dTIME measures the time between the current message and the last message with the same ID.
Delta time between messages is more useful than absolute timestamps considering that most CAN messages are periodic.
(iii) The R and T labels are mapped into 0 and 1 values. %
(iv) The various data logs are re-balanced by using random under-sampling in order to avoid training models biased towards the normal class, since the number of attack messages are heavily outnumbered by the number of normal messages.

After these steps, we obtain a dataset where each sample corresponds to a CAN message. The dataset in this format is used to train and test those ML models that get a single message input (\eg DNN). For the algorithms that require sequential inputs, we aggregate messages into frames with fixed size by using the sliding window method. The frames are labelled by considering the labels of the messages in the frame: if a frame includes 5 or more attack messages it is labelled as attack, otherwise it is considered as normal frame.

\subsection{Dataset splitting}
The pre-processed dataset is split into three subsets: \textit{Dataset A} used to train the IDS models,  \textit{Dataset B} used with the dual purpose of testing the IDS models and being perturbated for adversarial test, and \textit{Dataset C}   used with the dual purpose of testing baseline and retrained models and being perturbated for adversarial and defence tests.
We split the dataset randomly with a 60/20/20 ratio (where 60\% of the samples are allocated into Dataset A).
The distribution between normal and attack samples in all datasets is approximately equal. %

\subsection{IDS Training}
Four different ML-based IDSs are considered in this work.

\emph{BL-DNN} stands for BaseLine DNN model. The role of this model is to gain insights into the workings of IDS and provide a baseline regarding the performance for comparison between models. This model implements several techniques that prevent overfitting on the training data, in order to make it as robust as possible.

\emph{BL-Ensemble} stands for BaseLine Ensemble model. It is an ensemble model consisting of 5 different ML techniques. These techniques include Logistic Regression, Decision Tree, Support Vector Machine, K-Nearest Neighbors, and Naïve Bayes. Each of these techniques classifies an input sample independently, thus they vote for a certain class. The class with the most votes would then be returned as the output. The role of this model is to act as a second baseline regarding the performance. Additionally, it provides potential insights whether ensemble methods are more or less resilient towards adversarial samples.

\emph{SOTA-CNN} is initially introduced in \cite{In_Vehicle_Network_Intrusion_Detection_Deep_Convolutional_Neural_Network_2020} and uses a deep convolutional network architecture. The architecture of this model is a reduced version of the Inception-ResNet model proposed in \cite{inception_resnet_2017}. To use a CNN-based model for the CAN messages, a certain transformation has to be implemented. Song \etal \cite{In_Vehicle_Network_Intrusion_Detection_Deep_Convolutional_Neural_Network_2020} transforms several messages into frames. This is done by taking the message IDs of 29 consecutive messages and creating a matrix of these IDs. %
These frames of message IDs allowed their model to detect various attacks based on their ID and timing.

\emph{SOTA-LSTM} is based on an LSTM and is proposed in \cite{Anomaly_Detection_Automobile_Control_Network_Data_LSTM_2016}. It is divided into two parts. The first part includes the LSTM layers and aims to predict the next input in a sequence given the current input. This part is trained in an unsupervised manner, by only providing it with normal message data and letting it predict the desired next message. The goal is to minimize the difference between the predicted next input and the actual next input. The second part performs the actual classification. An input is given to the first part of the model, which then returns the predicted next input. The error between the predicted and actual next input is then compared to a certain threshold. If the error is above the threshold, the input is classified as an attack, otherwise the input is classified as a normal message. This threshold is set to the 99th percentile of the errors that originated from the training phase.

\subsection{Generation of adversarial data}

\begin{figure}[!ht]
  \begin{center}
    \includegraphics[scale=0.08]{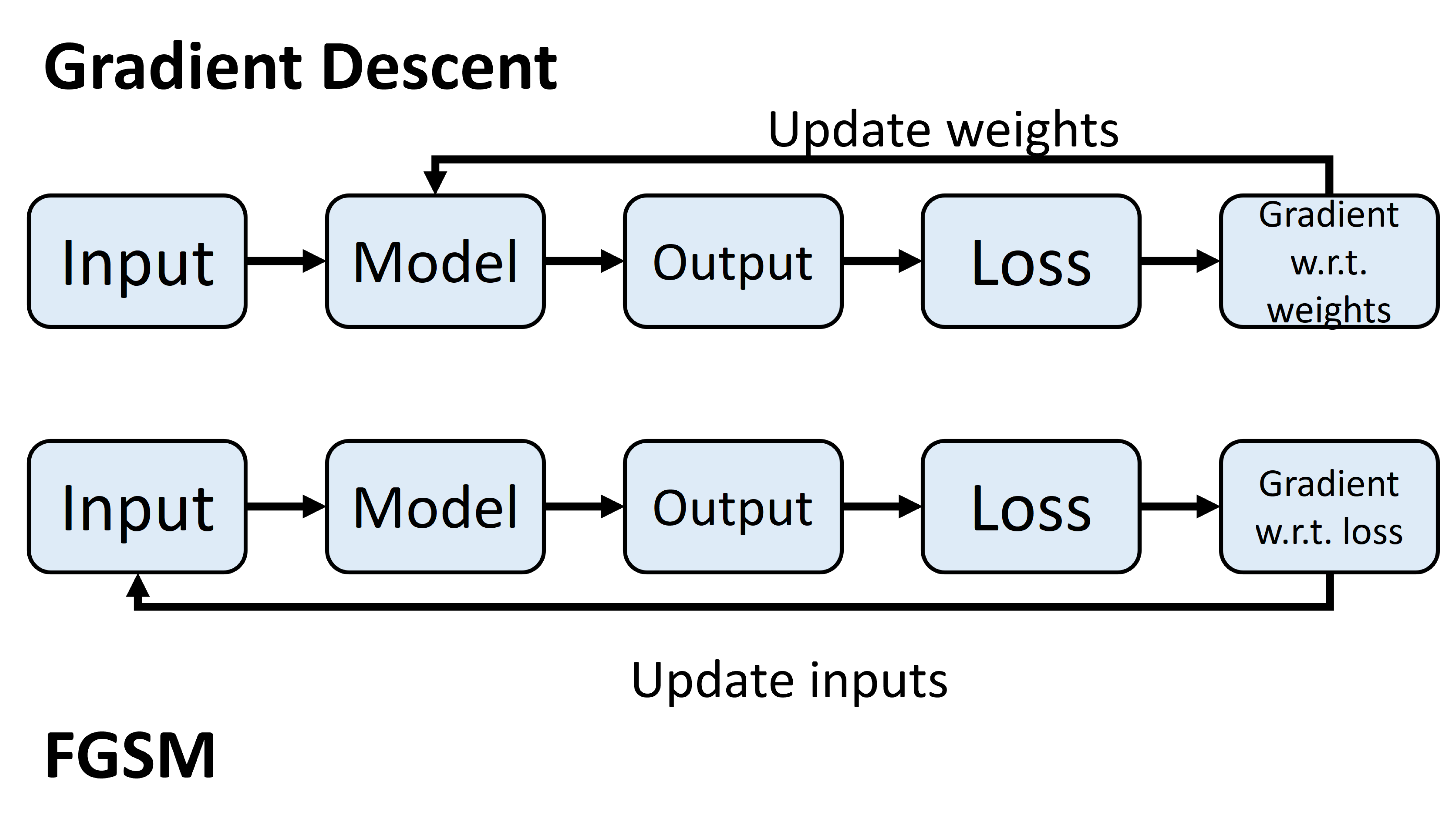}
  \end{center}
  \caption{Workflow of FGSM and gradient descent}
  \label{fig:fgsm-gradient-descent}
\end{figure}

Given the two \textit{Datasets B and C}, the Fast Gradient-Sign Method (FGSM) is used to generate adversarial samples. This results in two new datasets: \textit{$B^{\prime}$} and \textit{$C^{\prime}$}.
FGSM operates as follows (Figure~\ref{fig:fgsm-gradient-descent}): First, the input is fed into the model, then the output and the corresponding loss is computed. During gradient descent, the gradient of the loss is taken with respect to the model weights. This gradient is used to update the weights of the model to minimize the loss. For the FGSM, the first step is identical to gradient descent. However, the gradient of the loss is taken with respect to the input. This gradient is then used to compute the feature perturbation and update the inputs instead of the model weights. The perturbation can be formulated as:

\begin{equation}
    \eta = \epsilon * sign(\nabla_x J(\theta, x, y)),
\end{equation}

where the $J$ is the loss function for the model $\theta$. Once the gradient is computed, the sign of this gradient is calculated and multiplied with $\epsilon$, which is a small number that limits the amount of perturbation.

In order to adapt FGSM in the defined scenarios, we add two extensions based on the general idea of FGSM. %
The first extension provides control on which features to include during the adversarial sample generation so as to customize the dataset for attack scenario. On the other hand, second extension is an additional check to see if one of the input features has already reached the minimum or maximum value, and will be updated again in that direction according to the gradient. Since these updates will be reversed once the final step of clipping the values is reached, a certain loop might happen when the same feature is altered repeatedly. This is especially noticeable when using the L1-norm during the generation.

During the generation process, depending on what norm is used and what value for epsilon is taken, the success rate of the adversarial sample generation varies significantly. Whilst taking inspiration from the Basic Iterative Method \cite{Basic_Iterative_Method}, an additional loop function is implemented so that the FGSM runs until a successful perturbation is generated or until a maximum number of iterations. All the adversarial samples are generated on the BL-DNN model.

Although we consider a single perturbation algorithm, we define four different perturbation scenarios by varying the set of labels and features qualified for being altered. Thus, four adversarial datasets are obtained for a given input dataset.

\subsubsection{DoS adversarial scenario} only the last three ID bits and all data bits of DoS attack samples are perturbated. These constraints take into account the attacker limitation forced by the need of a successful DoS attack while evading the detection; in fact DoS attacks rely heavily on two factors: a low message ID (\eg a decimal ID of below 10) and a high transmission rate (\ie a low dTIME). For this reason, the dTIME feature is left out, and the message ID is capped at a maximum value of 7 by excluding all ID bits except for the final three ones.
\subsubsection{Fuzzy adversarial scenario} Only ID and data bits of Fuzzy attack samples are perturbated. This limitation is driven by the the main goal of the fuzzy attack that is to find combinations of message ID and data bits that result in unexpected behaviours.
\subsubsection{Malfunction adversarial scenario} The message ID bits, the DLC, and the dTIME are excluded for the perturbation of Malfunction attack samples. This is because the key aspects of these attacks are: (i) they target a specific message ID, (ii) messages are sent at specific intervals, and (iii) messages always include 8 data bytes.
\subsubsection{Full adversarial scenario} It considers the scenario that all the features for all the attacks in the input dataset are perturbated.

\subsection{Adversarial Training}
The adversarial training refers to the process of retraining a model with the adversarial samples to learn the adversarial patterns. By means of adversarial training, the model becomes more accustomed to adversarial samples and the underlying decision-boundary becomes robust towards such samples. According to the data flow shown in Figure~\ref{fig:data-flow}, we perform adversarial training on \textit{$Model_{A}$} with the \textit{dataset $B^{\prime}$}. By retraining the models on adversarial samples, the resulting models will be more resilient towards these adversarial samples.

\subsection{Testing}
This paper covers three different testing experiments. As shown in Figure~\ref{fig:data-flow}, firstly the \textit{Baseline test}, where the original models \textit{$Model_{A}$} trained on \textit{Dataset A} are tested on \textit{Dataset B} and \textit{C}. This test is to set a baseline regarding the performance of the models on the non-adversarial datasets.

Second, in the \textit{Adversarial test}  the original models \textit{$Model_{A}$} are tested on \textit{Dataset $B^{\prime}$} and \textit{$C^{\prime}$}. %
The resulting performance provides insights into the effectiveness of the adversarial samples in evading detection as well as the transferability of these samples. The transferability is visible in the results of the BL-Ensemble, SOTA-CNN and SOTA-LSTM models, since the adversarial samples are all generated by exploiting the BL-DNN model.

Finally,  the original models \textit{$Model_{A}$} that underwent adversarial training are tested on \textit{Dataset C} and \textit{$C^{\prime}$} in the \textit{Defence test}. The results of these experiments aim to show that (i) adversarial samples can be detected if we know the algorithm used for data perturbation so that the models can be retrained with perturbed samples, and at the same time (ii) the performance on the non-adversarial samples are preserved.

\section{Implementation}
\label{sec:evaluation}

To evaluate the effect of adversarial samples and adversarial training, a large number of experiments are conducted. The FGSM used in these experiments is based on the source code provided by CleverHans.\footnote{https://github.com/cleverhans-lab/cleverhans} All models are implemented in python with Tensorflow.\footnote{https://www.tensorflow.org/}

\subsection{Metrics}

For the performance evaluation we consider the following metrics: Accuracy, False Negative Rate (FNR), False Positive Rate (FPR) and F1-score. Note that in presence of unbalanced data (which is very common in security datasets), accuracy would be somewhat misleading. F1-score overcomes this issue by computing the harmonic mean of the precision and recall. For example, in a $80/20$ dataset a model that never detects any attack would  have $80\%$ accuracy but an F1-score of $0$.

FNR and FPR are used to clarify what type of errors are made by the model. In particular, a successful evasion attack is evident by a high FNR. Having near-to-zero FPR is fundamental, especially in IDSs for automotive networks because of the missing ability in remote handling of network incidents, while automated handling of alerts may result in triggering useless and massive out-of-service issues.

\subsection{Models}
Each of the considered models requires a customized fine-tuning process in order to take design decisions and select the values of some parameters so as to calibrate the outputs. %

\subsubsection{BL-DNN}
The model consists of 3 dense layers with 128, 64, and 2 neurons respectively. These layers are separated by a normalization layer, which normalizes the intermediate outputs. Additionally, the first two dense layers use the ReLu activation function and applies the L2 weight regularizar. The regularizer prevents the weights of the neurons from getting too large of a magnitude. The final dense layer uses the softmax activation function to perform the final classification. All the weights of the model are initialized using TensorFlow's default initializer. The model uses the Adam optimizer with a learning\_rate of 0.001 and has the Sparse Categorical Crossentropy as the loss function.

\subsubsection{BL-Ensemble}
The ensemble method consists of 5 different ML techniques including of logistic regression, decision tree, support vector machine, K-nearest neighbor, and Gaussian Naïve Bayes. These 5 sub-models are combined into a voting classifier which uses hard voting. This means that the label predictions of each sub-model are collected, and the majority vote becomes the final predicted label.

\subsubsection{SOTA-CNN}
This model consists of the following parts: a Stem block, a Inception-ResNet-A block, a Reduction-A block, Inception-ResNet-B block, a Reduction-B block, an Average Pool layer, a Dropout layer, and a final dense softmax layer. The intermediate results are fed into the last pooling, dropout and dense layer. The model uses the Adam optimizer with a learning\_rate of 0.001 and the sparse categorical cross-entropy as the loss function. It is trained for 10 epochs, with the option of early stopping if the validation accuracy does not improve for 3 epochs.

\begin{figure*}[!ht]
  \begin{center}
    \includegraphics[width=\textwidth]{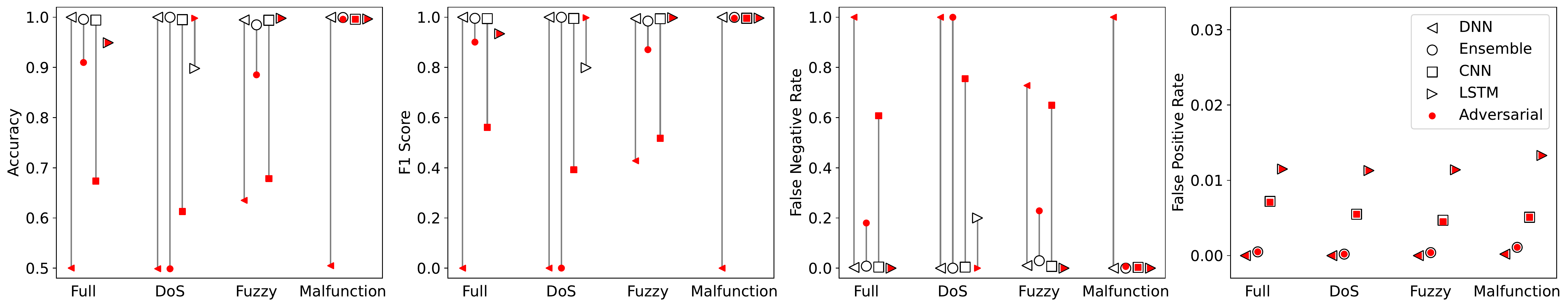}
  \end{center}
  \caption{Metrics of standard models with and without adversarial samples}
  \label{fig:standard-metrics}
\end{figure*}

\subsubsection{SOTA-LSTM}
Each SOTA-LSTM model is trained on messages of a single ID and has the following architecture and parameter settings. The models first have two dense layers with 128 neurons each and the hyperbolic tangent as activation function. These layers are followed by two LSTM layers of 512 neurons each, which also use the hyperbolic tangent as activation function. Afterwards, there is a final dense layer with 64 neurons and the Sigmoid activation function. All the layers described before are alternated with a dropout layer and dropout rate of 0.2. All layer weights are initialized in the same way as the BL-DNN model. The models use the Adam optimizer with a learning\_rate of 0.001 and the binary crossentropy as the loss function. The model is trained for 10 epochs, but early stopping is implemented if the validation loss does not improve for 3 epochs.

\subsubsection{FGSM}
All adversarial samples are generated on the BL-DNN model. The norm and epsilon values used are L1 and 1.0, respectively. This ensures that only one of the features is altered in every iteration with a step size of 1.0. The minimum value of all input features is set to 0, to prevent negative values. And the maximum value of all bit features (\ie the ID bits and the 64 Data bits) is set to 1. The maximum value for the DLC and dTIME features is set to 8 and 100, respectively. Finally, the maximum number of iterations is set to 50, and the target for the adversarial samples is set to 0 (such that the attacks are classified as normal messages).

\section{Results and Discussion}
\label{sec:result}
First, we analyze the performance on each selected model tested against regular and adversarial samples. Second, we study the effectiveness of adversarial training. Third, we look into the details of adversarial samples. Finally, we study how different features perform in the adversarial sample generation.

\subsection{Analysis of Overall Performance}

The results of the different models before retraining can be seen in Figure~\ref{fig:standard-metrics} where the four subplots correspond to the four metrics used: accuracy, F1-score, FNR, FPR. The metrics are recorded for each model on each attack scenario. Different markers are used to represent different models; also the markers with a white backgrounds refer to the performance obtained by \textit{Dataset B + C}, and the markers with red backgrounds refer to adversarial samples in \textit{Dataset B' + C'}. %

\begin{tcolorbox}[float=t,
                  colback=gray!30,%gray background
                  colframe=black,% black frame colour
                  width=\columnwidth,% Use 5cm total width,
                  boxrule=0.5pt,
                  arc=1mm, auto outer arc,
                 ]
\noindent\textbf{Takeaway A1:} Performance of almost all the ML-based IDSs has been significantly impacted by adversarial samples, except for SOTA-LSTM.

\end{tcolorbox}

Overall, we can observe from  Figure~\ref{fig:standard-metrics} that three out of  four models are vulnerable and sensitive to the adversarial samples. Among of all these three models, BL-DNN performs the best in the normal situation, but at the same time, it is impacted the most by the adversarial samples. This large drop in performance is to be expected, since the perturbations are generated on the BL-DNN. On the other hand, the performance of SOTA-CNN on Malfunction dataset is not affected by adversarial samples. It is because only data bits are flipped over in the Malfunction dataset, while SOTA-CNN only uses the ID bits as inputs, meaning that there is no difference between the regular and the adversarial samples in this scenario. This also explains why SOTA-CNN performs the worst on the DoS adversarial dataset.

\begin{figure*}[!ht]
  \begin{center}
    \includegraphics[width=\textwidth]{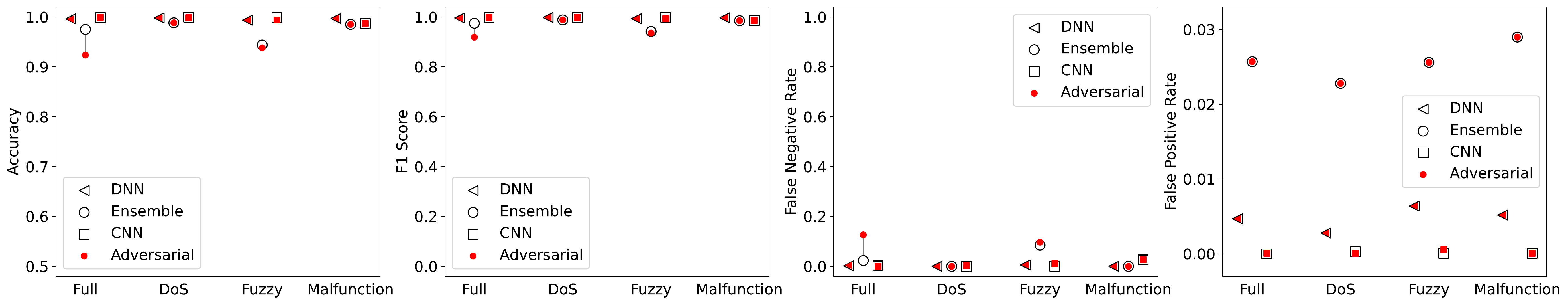}
  \end{center}
  \caption{Metrics of retrained models with and without adversarial samples}
  \label{fig:retrained-metrics}
\end{figure*}

It is important to note that SOTA-LSTM performs noticeably the worst for Full and DoS datasets when the data are clean. However, the results show that the adversarial samples are not able to evade detection of SOTA-LSTM, while, the performance on the DoS adversarial dataset even improves once the adversarial samples are added.
The possible reason to this observation as to the limited influence of the perturbations lies in the nature of recurrent neural networks. This type of networks (\eg SOTA-LSTM) not only learns from the data within an input, but also from the correlation among bits within a frame. The current adversarial samples are generated by BL-DNN, meaning that only the information within a stand-alone message is altered by the perturbations. But the information associated to the attack and lying on correlation between messages is still present, and it is enough for the  SOTA-LSTM to correctly classify the adversarial samples.

\begin{tcolorbox}[float=t,
                  colback=gray!30,%gray background
                  colframe=black,% black frame colour
                  width=\columnwidth,% Use 5cm total width,
                  boxrule=0.5pt,
                  arc=1mm, auto outer arc,
                 ]
\noindent\textbf{Takeaway A2:} The results show transferability; in fact adversarial samples generated on the BL-DNN model impact the performance of other ML-based IDSs, especially in the scenario of DoS attack.
\end{tcolorbox}

As expected, no significant difference can be observed in FPRs, since adversarial samples aim to affect only intrusion detection rate.
Note that
the drop in performance not only signifies a susceptibility to adversarial samples, but also the transferability of these samples. In fact, while the adversarial samples are created by BL-DNN, they are able also to evade the detection of other ML models. From the results we can see that the adversarial samples in the DoS adversarial dataset are especially transferable, while the adversarial samples in the Malfunction dataset have no effect on the other models.

\subsection{Performance Analysis of Adversarial Training}

\begin{tcolorbox}[colback=gray!30,%gray background
                  colframe=black,% black frame colour
                  width=\columnwidth,% Use 5cm total width,
                  boxrule=0.5pt,
                  arc=1mm, auto outer arc,
                 ]
\noindent\textbf{Takeaway B:} The impact of the adversarial samples is nearly negligible when adversarial training is applied to the models.

\end{tcolorbox}

In this experiment, we investigate the effeteness of adversarial training as a state-of-the-art defense technique on the compromised datasets. By retraining the previously-mentioned models on the adversarial samples, they become more resistant to these samples. As mentioned earlier, the models are retrained on \textit{Dataset B'} and tested on \textit{Dataset C + C'}. Note that the \textit{SOTA-LSTM} is not retrained and tested since the majority of the adversarial samples do not influence their performance beforehand, suggesting that no noticeable results would appear after retraining. The results of the adversarial-trained models on the four datasets with and without adversarial samples are illustrated in Figure~\ref{fig:retrained-metrics}.
In general,  BL-DNN's performance slightly drops, as there is a small FPR increase.
Also, the performance of the BL-Ensemble on the non-adversarial datasets has slightly decreased after retraining, especially on the Full and Fuzzy datasets.

\subsection{Analysis of Adversarial Samples}

\begin{tcolorbox}[colback=gray!30,%gray background
                  colframe=black,% black frame colour
                  width=\columnwidth,% Use 5cm total width,
                  boxrule=0.5pt,
                  arc=1mm, auto outer arc,
                 ]
\noindent\textbf{Takeaway C:} A successful adversarial perturbation can be obtained within a few iterations of the adversarial algorithm, and it would also require only few features to be modified.
\end{tcolorbox}

\begin{figure}[ht]
  \begin{center}
    \includegraphics[scale=0.42]{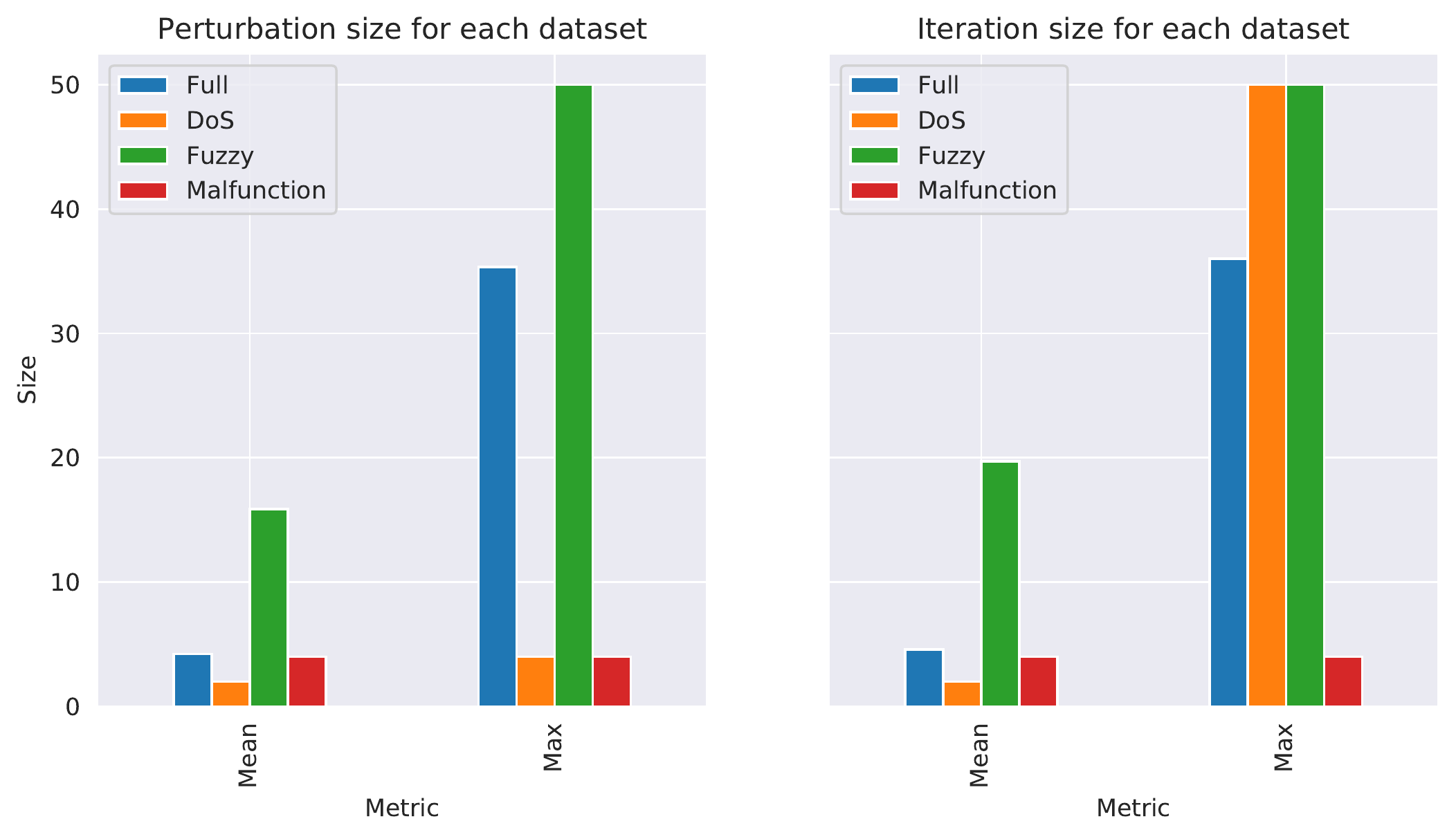}
  \end{center}
  \caption{Statistics of perturbation size and iterations}
  \label{fig:pertubation_iteration}
\end{figure}

In this experiment, we look at the statistics of the perturbation size, \ie the number of features that are modified for the generation of adversarial samples, and the number of iterations required by the adversarial generation algorithm to generate a sample in the four different scenarios (Full, DoS, Fuzzy, and Malfunction adversarial). Figure~\ref{fig:pertubation_iteration} shows the mean and max metrics for both perturbation size and number of iterations per scenario. Note that the figure also shows the statistics of the adversarial samples that are not generated successfully. This means that the samples that are not generated within 50 iterations are included in these statistics.

Figure~\ref{fig:pertubation_iteration} shows the statistical results of the perturbation size and iterations for the adversarial generation algorithm. Overall, if the adversarial generation algorithm converges, it generates a sample in a relatively small number of iterations and features. Especially in DoS adversarial scenario, there are only 2 iterations required and 2 features are modified to generate adversarial samples. Fuzzy adversarial scenario, on the other hand, provided the largest amount of failed samples.

\begin{figure*}[!ht]
    \begin{center}
    \includegraphics[scale=0.48]{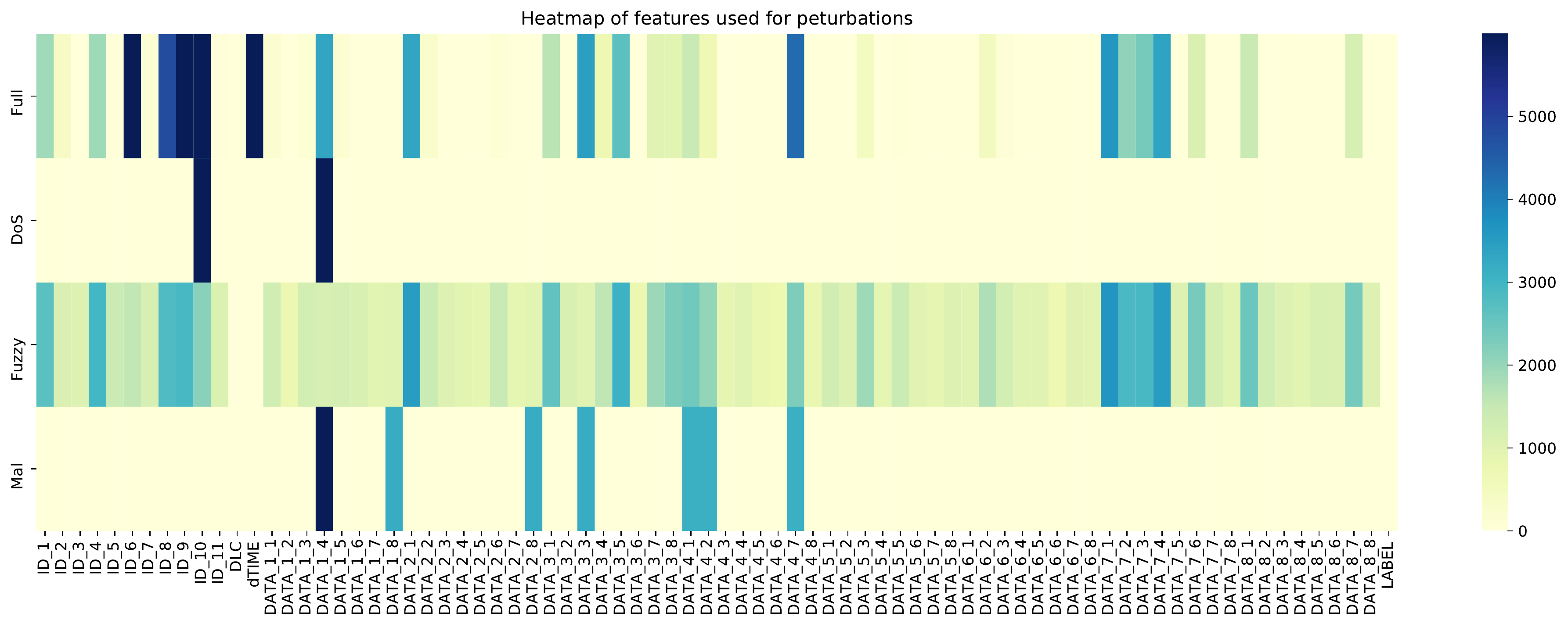}
    \end{center}
    \caption{Heatmap of the perturbation features}
    \label{fig:Pert_heat}
\end{figure*}

\subsection{Feature Analysis}

\begin{tcolorbox}[colback=gray!30,%gray background
                  colframe=black,% black frame colour
                  width=\columnwidth,% Use 5cm total width,
                  boxrule=0.5pt,
                  arc=1mm, auto outer arc,
                 ]
\noindent\textbf{Takeaway D:} Only some specific features, especially those with the lowest ID bits, are the most relevant to the generation of adversarial samples.

\end{tcolorbox}
Now, a closer look is taken to understand which features played the largest role in the adversarial sample generation. Figure~\ref{fig:Pert_heat} shows a heatmap for the four adversarial datasets so that the prominence of each input feature is emphasized.
In the heatmap, various interesting patterns are visible. Firstly, the adversarial sample generation for both  Full and  Fuzzy adversarial scenarios determines a high number of features to be altered. However,  Full adversarial dataset seems to put larger priority on several message ID bits. In contrast to the Full and Fuzzy dataset, both DoS and Malfunction datasets only include a limited number of perturbed features. For the DoS adversarial dataset, only two features mainly get perturbed. These features are the 10th message ID bit and the 4th data bit. The Malfunction dataset  includes only 7 perturbed features with the 4th data bit being the one perturbed most frequently.
In general, the lowest ID bits seem  to be quite influential in the adversarial sample generation. This could signal that the models highly depend on these bits, especially when you look at the performance obtained on the DoS dataset. The 4th data bit additionally plays a large role both in the DoS and  Malfunction datasets. Finally, there seems to be a small hotspot around the first four bits of the 7th data byte. These bits are more relevant for the Full and Fuzzy datasets. Doing further feature analysis is outside the scope of the paper, so currently no reason can be provided as to why these features are so relevant.

\section{Conclusion and Future work}
\label{sec:conclusion}

In this paper, we have pointed out the limitations and weaknesses present in certain ML- or DL-based IDSs by evaluating the resilience of selected models for automotive networks and demonstrated their vulnerability to adversarial samples. Furthermore, we have investigated transferability of the adversarial samples and demonstrated that even when adversarial samples are generated from a simplified base model, they are (with some limitations) still able to evade the target IDS. This demonstrates that an attacker does not need to know the IDS implementation in order to launch effective attacks. Interestingly, algorithms trained on sequential CAN messages are successful in detecting these evasion attacks as they are capable of capturing in their models the correlations within a sequence of messages.

Our future work will focus on: (i) analysing  different base models for the adversarial sample generation, (ii) extending the generation to adversarial sequences and analyzing its limitations, and (iii) investigating how the adversarial sample generation can be adapted to evade detection of retrained models.

\section{Acknowledgements}
This research is partially funded by the CyReV project (Sweden's Innovation Agency, D-nr 2019-03071), partially by the H2020  ARCADIAN-IoT (Grant ID. 101020259), and  H2020 VEDLIoT (Grant ID. 957197).

\bibliographystyle{IEEEtran}
\bibliography{references}

\end{document}